# Temporal rise in the proportion of younger adults and older adolescents among COVID-19 cases in Germany: evidence of lesser adherence to social distancing practices?

Edward Goldstein[1,*]

1) Center for Communicable Disease Dynamics, Department of Epidemiology, Harvard T.H. Chan School of Public Health, 677 Huntington Ave., Boston, MA 02115, USA
*) Corresponding author, egoldste@hsph.harvard.edu

**Abstract**

Background: There is a great deal of uncertainty about the role of different age groups in propagating the SARS-CoV-2 epidemics in different countries, particularly under the current social distancing practices.

Methods: We used the Robert Koch Institute data on weekly COVID-19 cases in different age groups in Germany. To minimize the effect of changes in healthcare seeking behavior and testing practices, we included the following age groups in the analyses: 10-14y, 15-19y, 20-24y, 25-29y, 30-34y, 35-39y, 40-44y, 45-49y. For each age group $g$ above, we considered the proportion $PL(g)$ of individuals in age group $g$ among all detected cases aged 10-49y during weeks 13-14, 2020 (later period), as well as corresponding proportion $PE(g)$ for weeks 10-11, 2020 (early period), and defined the relative risk $RR(g)$ for the age group $g$ to be the ratio $RR(g) = PL(g)/PE(g)$. For each pair of age groups $g1, g2$, a higher value of $RR(g1)$ compared to $RR(g2)$, or, alternatively, a value above 1 for the odds ratio $OR(g1, g2) = RR(g1)/RR(g2)$ for a COVID-19 case to be in group $g1$ vs. $g2$ for the later vs. early periods is interpreted as the relative increase in the population incidence of SARS-Cov-2 in the age group $g1$ compared to $g2$ for the later vs. early period.

Results: The relative risk $RR(g)$ was highest for individuals aged 20-24y (RR=1.4(95% CI (1.27,1.55))), followed by individuals aged 15-19y (RR=1.14(0.99,1.32)), aged 30-34y (RR=

1.07(0.99,1.16)), aged 25-29y (RR= 1.06(0.98,1.15)), aged 35-39y (RR=0.95(0.87,1.03)), aged 40-44y (RR=0.9(0.83,0.98)), aged 45-49y (RR=0.83(0.77,0.89)) and aged 10-14y (RR=0.78(0.64,0.95)). For the age group 20-24y, the odds ratio relative to any other age group for a case to be during the later vs. early period was significantly above 1. For the age group 15-19y, the odds ratio relative to any other age group either above 35y or 10-14y for a case to be during the later vs. early period was significantly above 1.

Conclusions: The observed relative increase with time in the prevalence of individuals aged 15-34y (particularly those aged 20-24y) among detected COVID-19 cases in Germany is unlikely to be explained by increases in the likelihood of seeking medical care or the likelihood of being tested for individuals in those age groups compared to individuals aged 35-49y or 10-14y, and should be indicative of the actual increase in the prevalence of individuals aged 15-34y among SARS-CoV-2 infections in the German population. That increase likely reflects elevated mixing among individuals aged 15-34y (particularly those aged 20-24y) compared to other age groups, possibly due to lesser adherence to social distancing practices.

**Introduction**

The ongoing SARS-CoV-2 pandemic has caused over 1,133,000 detected cases of COVID-19 illness worldwide and claimed over 62,700 lives as of April 5, 2020 [1]. Various forms of social distancing measures/testing practices were implemented in different countries/regions in order to stem the spread of the epidemic. Under those social distancing measures, rates of contact between individuals in different age groups are expected to significantly depart from the regular mixing patterns [2]. Under such circumstances, there is a great deal of uncertainty regarding the role of different age groups in propagating the SARS-CoV-2 epidemics in different countries. While disease is most severe in older age groups, a sizeable share of COVID-19 related hospitalizations in Western countries belongs to individuals aged 20-55y [3]. A study of close contacts of COVID-19 cases in China [4] found comparable rates of infection with SARS-CoV-2 in different age groups. A better understanding is needed of the effect that different age groups play in propagating the SARS-CoV-2 pandemic during its current stage.

In this paper, we apply the methodology in [5,6] to assess the relative roles of different age groups during the early stage of the 2020 SARS-CoV-2 epidemic in Germany using data on COVID-19 cases provided by the Robert Koch Institute [7]. The idea of that method is that age groups that have an elevated role in propagating an epidemic due to higher contact rates and/or susceptibility to infection will have their share among all incident cases increase with time during the early states of the epidemic, particularly under containment efforts for which adherence varies between the different age groups. For example, if contact rates in one age group increase with time relative to another due to lesser adherence, incidence of infection in the former age group will increase relative to the latter. Additionally, if a given country comprises several regions with different growth rates for their epidemics, the age group that drives the incidence of infection in regions with faster growing/larger epidemics will have its share among all incident cases in the country increase with time. Temporal increase in the share of a given age group among all cases of infection can be evaluated using the relative risk (RR) statistic that estimates the ratio of the proportion of a given age group among all detected (reported) cases of COVID-19 for a later time period vs. an earlier time period. We note that for a given age group, the ratio between the number of detected COVID-19 cases and incident cases of SARS-CoV-2 infection (case-detection rate) in that age group may vary with time due to changes in healthcare-seeking behavior and other factors. To minimize the effect of relative changes, for the different age groups, in the case-detection rates with time, we restrict our analysis to individuals aged 10-49y only as symptomatic older adults and younger children may be more likely to seek medical care/get tested compared to other age groups as the awareness of the dangers of the epidemic increased with time.

**Methods**

*Data*

Data on COVID-19 cases in Germany stratified by week and 5-year age groups can be accessed through the Robert Koch Institute via the SurvStat@RKI 2.0 application [7]. Those data can be extracted by creating a query, selecting the attribute to be Disease/Pathogen, selecting the disease to be Covid-19, and the attributes to display to be 5-year age groups and week of notification.

*Statistical analysis*

We select the early period to be weeks 10-11, 2020, and the later period to be weeks 13-14, 2020. Note that at the time of data access (Apr. 4, 2020), data for week 14, 2020 were incomplete. We include the following eight age groups in our analysis: 10-14y, 15-19y, 20-24y, 25-29y, 30-34y, 35-39y, 40-44y, 45-49y. For each age group $g$, let $E(g)$ be the number of detected COVID-19 cases in age group $g$ during the early period, and $L(g)$ be the corresponding number during the later period. The relative risk statistic is

$$RR(g) = \frac{L(g)}{\sum_{h=1}^{8} L(h)} \bigg/ \frac{E(g)}{\sum_{h=1}^{8} E(h)} \qquad (1)$$

The observed numbers of detected cases $L(g)$ and $E(g)$ in the age group $g$ during the later and early periods are binomially distributed (with the total equaling $\sum_{h=1}^{8} L(h)$ and $\sum_{h=1}^{8} E(h)$ correspondingly). Moreover, we assume that the numbers of reported cases are sufficiently high so that the logarithm $\ln(RR(g))$ of the relative risk in the age group $g$ is approximately normally distributed [8]. Under this approximation, the 95% confidence interval for $RR(g)$ is $exp(\ln(RR(g)) \pm 1.96 \cdot SE)$, where $\ln(RR(g))$ is estimated via eq. 1, and the standard error is

$$SE = \sqrt{\frac{1}{L(g)} + \frac{1}{E(g)} - \left(\frac{1}{\sum_{h=1}^{8} L(h)} + \frac{1}{\sum_{h=1}^{8} E(h)}\right)} \qquad (2)$$

For each pair of age groups $g1, g2$, comparison of the relative risks $RR(g1)$ and $RR(g2)$ is performed using the odds ratio

$$OR(g1, g2) = RR(g1)/RR(g2) \qquad (3)$$

We note that it follows immediately from eq. 1 that $OR(g1, g2)$ equals the odds ratio $\frac{L(g1)}{E(g1)} \big/ \frac{L(g2)}{E(g2)}$ for a COVID-19 case to be in age group $g1$ vs. $g2$ for the later vs. early period.

Estimates and confidence intervals for pairwise odds ratios are performed using Fisher's exact test.

**Results**

Table 1 shows the estimates for the relative risk $RR(g)$ for being in a given age group for a detected COVID-19 case during the later period (weeks 13-14, 2020) vs. early period (weeks 10-11, 2020) for each of the eight age groups used in our analysis. The highest estimate for the relative risk belongs to individuals aged 20-24y, followed by individuals aged 15-19y, 30-34y and 25-29y, with estimates of the relative risk for individuals aged over 35y and those aged 10-14y being lower.

| Age group | RR (relative risk) |
| --- | --- |
| 10-14y | 0.78(0.64,0.95) |
| 15-19y | 1.14(0.99,1.32) |
| 20-24y | 1.4(1.27,1.55) |
| 25-29y | 1.06(0.98,1.15) |
| 30-34y | 1.07(0.99,1.16) |
| 35-39y | 0.95(0.87,1.03) |
| 40-44y | 0.9(0.83,0.98) |
| 45-49y | 0.83(0.77,0.89) |

**Table 1:** Relative risks for being in a given age group for COVID-19 cases during the late period (weeks 13-14) vs. early period (weeks 10-11) for each of the eight age groups included in our analysis. Note that the data for week 14 are incomplete and extracted from [7] on 04/04/2020.

Table 2 shows, for different pairs of age groups, the estimated odds ratios for being a detected COVID-19 case during the later vs. early period for one age groups vs. the other. For the age group 20-24y, the odds ratio relative to any other age group for a case to be during the later vs. early period was significantly above 1. For the age group 15-19y, the odds ratio relative to any other age group either above 35y or 10-14y for a case to be during the later vs. early period was significantly above 1. For the age groups 25-29y and 30-34y, the odds ratio relative to any other age group either above 40y or 10-14y for a case to be during the later vs. early period was significantly above 1.

| Age group | 15-19y | 20-24y | 25-29y | 30-34y | 35-39y | 40-44y | 45-49y |
|---|---|---|---|---|---|---|---|
| 10-14y | **0.68 (0.53,0.89)** | **0.56 (0.44,0.71)** | **0.74 (0.59,0.93)** | **0.73 (0.58,0.92)** | 0.82 (0.66,1.04) | 0.86 (0.69,1.09) | 0.94 (0.76,1.18) |
| 15-19y | | **0.81 (0.68,0.98)** | 1.08 (0.9,1.29) | 1.07 (0.9,1.28) | **1.21 (1.01,1.44)** | **1.26 (1.06,1.51)** | **1.38 (1.17,1.64)** |
| 20-24y | | | **1.32 (1.15,1.53)** | **1.31 (1.14,1.51)** | **1.48 (1.29,1.71)** | **1.55 (1.35,1.79)** | **1.7 (1.48,1.94)** |
| 25-29y | | | | 0.99 (0.87,1.13) | 1.12 (0.99,1.27) | **1.17 (1.03,1.33)** | **1.28 (1.14,1.44)** |
| 30-34y | | | | | 1.13 (0.99,1.29) | **1.18 (1.04,1.35)** | **1.29 (1.15,1.46)** |
| 35-39y | | | | | | 1.05 (0.92,1.19) | **1.14 (1.02,1.29)** |
| 40-44y | | | | | | | 1.09 (0.97,1.23) |

**Table 2:** Odds ratios, for different pair of age groups, for a COVID-19 case to be in the later period (weeks 13-14) vs. early period (weeks 10-11) for one age group vs. the other. Note that the data for week 14 are incomplete and extracted from [7] on 04/04/2020.

**Discussion**

There is a great deal of uncertainty about the role of different age groups in propagating the ongoing COVID-19 epidemics in different countries, particularly in light of the current social distancing measures and testing practices. Some evidence about a larger relative role for certain age groups vs. other can be obtained by examining temporal changes in the proportions of different age groups among detected COVID-19 cases, as explained in the 2nd paragraph of the Introduction. This estimation is done using the relative risk (RR) statistic that we have employed in our previous work [5,6]. Our results, based on applying this method to data on COVID-19 cases in Germany provided by the Robert Koch Institute [7], suggest that individuals aged 15-34y (particularly those aged 20-24y) have a greater relative role in propagating the SARS-CoV-2 epidemic in Germany compared to individuals aged 35-49y and children aged 10-14y. Some of those differences in relative roles during the ongoing epidemic may stem from higher mixing related to lesser adherence to social distancing guidelines for individuals aged 15-34y (particularly those aged 20-24y). Further work is needed to assess those issues in different countries.

Our paper has some limitations. One limitation is that the ratio between the number of detected COVID-19 cases and incident cases of SARS-CoV-2 cases (case-detection rate) may vary with time in each age group. We believe that it is unlikely that this phenomenon is more pronounced for individuals aged 15-34y compared to persons aged 35-49, namely that increases in the likelihood of seeking medical care or the likelihood of being tested for individuals aged 15-34y are substantially larger compared to individuals aged 35-49y or 10-14y. Therefor, the observed temporal increases in the relative share of individuals aged 15-34y among detected COVID-19 cases in persons aged 10-49y should be indicative of the actual increase in the prevalence of individuals aged 15-34y among SARS-CoV-2 infections in the German population. Another limitation is the uncertainty regarding the relation between the temporal rise in the share of a given age group among SARS-CoV-2 cases and the role that this age group plays in propagating the SARS-CoV-2 epidemic. Some of the factors supporting that relation are described in the 2nd paragraph of the Introduction. Additionally, data on mixing patterns in different age groups during regular times [2], as well as some evidence about the current social interactions of younger adults/older adolescents suggests that those age groups are expected to have the largest role in driving the current SARS-CoV-2 outbreaks, which is indeed supported by our results.

We believe that despite those limitations, our results provide evidence about the role of younger adults (particularly those aged 20-24y) and older adolescents in driving the current COVID-19 epidemic in Germany, with that role potentially being related to lesser adherence to social distancing guidelines. Those results may be relevant to informing social distancing efforts, particularly for younger adults and older adolescents.